\documentclass{svjour3}                     
\smartqed  
\usepackage{graphicx}
\usepackage{mathptmx}      
\usepackage{latexsym}
\journalname{International Journal of Theoretical Physics}

\begin{document}

\title{Entanglement and Berry Phase in a $(3\times 3)-$dimensional Yang-Baxter system}


\author{Gangcheng Wang \and  Chunfang Sun \and  Qingyong Wang \and Kang
Xue}

\institute{Gangcheng Wang \at
              School of Physics, Northeast Normal University,
Changchun 130024, People's Republic of China
              \email{wanggc887@nenu.edu.cn}           
           \and
           Chunfang Sun \at
              School of Physics, Northeast Normal University,
Changchun 130024, People's Republic of China \and  Qingyong Wang \at
              School of Physics, Northeast Normal University,
Changchun 130024, People's Republic of China \and Kang Xue\at
              School of Physics, Northeast Normal University,
Changchun 130024, People's Republic of China
\email{XueKang@nenu.edu.cn}}

\date{Received: date / Accepted: date}

\maketitle

\begin{abstract}
Based on the method which is given in Ref. [Sun et.al.
arXiv:0904.0092v1], we present another $9\times 9$ unitary
$\breve{R}-$matrix, solution of the Yang-Baxter Equation, is
obtained in this paper. The entanglement properties of
$\breve{R}-$matrix is investigated, and the arbitrary degree of
entanglement for two-qutrit entangled states can be generated via
$\breve{R}$-matrix acting on the standard basis. A Yang-Baxter
Hamiltonian can be constructed from unitary $\breve{R}-$matrix. Then
the geometric properties of this system is studied. The results
showed that the Berry phase of this system can be represented under
the framework of SU(2) algebra. 
\end{abstract}
\section{introduction}
\label{sec1}

Quantum entanglement(QE), the most surprising nonclassical property
of quantum systems, plays a key role in quantum information and
quantum computation processing\cite{Bennett,ben1,ben2,mdmv}. Because
of these applications, QE has become one of the most fascinating
topics in quantum information and quantum computation. On the other
hand, the geometrical phase\cite{berry}, such as Berry phase(BP), is
another important concept in quantum
mechanics\cite{aa,Esj,JR,DELC,FA}. In recent years, a lot of works
have been attributed to BP\cite{AAHQJZ}, because of its possible
applications to quantum computation(the so-called geometric quantum
computation)\cite{J. Jones,W. K.,A. Ekert}. Such concern is
motivated by the belief that geometric quantum gates should exhibit
an intrinsic fault tolerance in the presence of some kind of
external noise due to the geometric nature of the BP.

  Yang-Baxter Equation(YBE)\cite{yang,baxter,drin} was originated in solving quantum
  integrable models, but recently has been shown to have a deep
  connection with topological quantum computation and entanglement
  swapping\cite{qiybe1,kauffman1,qiybe3,zkg,zg,ckg,ckg2,cxg1}. In Ref.\cite{hgx} , the authors point out YBE can be tested in
  terms of quantum optics. In a very recently work\cite{cxg1}, it is found
  that any pure two-qudit entangled state can be achieved by a
  universal Yang-Baxter Matrix assisted by local unitary
  transformations. However, the solution $\breve{R}(x)-$matrix in
  Ref.\cite{cxg1} only dependent on one parameter. So we can't construct a
  Yang-Baxter Hamiltonian as in Ref.\cite{ckg,sun}. In this paper, we obtain
  a time-dependent
  solution of YBE, $\breve{R}(x,\varphi_{1},\varphi_{2})$.
  $\varphi_{1}$ and $\varphi_{2}$ are time-dependent, so we can
  construct Yang-Baxter Hamiltonian. Consequently, we can study
  entanglement properties and Berry phase for this system.

  This paper is organized as follows: In Sec\ref{sec2}, we
  present a $9\times 9$ Yang-Baxter matrix. By means of negativity,
  we investigated the entanglement properties of
  $\breve{R}(x,\varphi_{1},\varphi_{2})$-matrix. We show that the arbitrary degree of
entanglement for two-qutrit entangled states can be generated via
the unitary matrix
$\breve{R}(\theta,\varphi_{1},\varphi_{2})$-matrix acting on the
standard basis. In Sec\ref{sec3}, we construct a Hamiltonian from
the unitary $\breve{R}(\theta,\varphi_{1},\varphi_{2})$-matrix. The
Berry phase of the system is investigated, and the results showed
that the Berry phase of this system can be represented under the
framework of SU(2) algebra. The summary is made in the last section.

\section{Unitary solution of Yang-Baxter Equation and its
entanglement properties}\label{sec2}

 The usual YBE takes the form,
\begin{equation}\label{ybe}
\breve{R}_{i}(x)\breve{R}_{i+1}(xy)\breve{R}_{i}(y)=\breve{R}_{i+1}(y)\breve{%
R}_{i}(xy)\breve{R}_{i+1}(x)
\end{equation}
The spectral parameters x and y which are related with the
one-dimensional momentum play an important role in some typical
models\cite{yang}. The asymptotic behavior of
$\breve{R}(x,\varphi_{1},\varphi_{2})$ is x-independent, \emph{i.e.}
$lim\breve{R}_{i,i+1}(x,\varphi_{1},\varphi_{2})\propto b_{i}$,
where $b_{i}$ are braiding operators, which satisfy the braiding
relations,
\begin{eqnarray}\label{braid}
\left\{
\begin{array}
[c]{ll}
b_{i}b_{i+1}b_{i}=b_{i+1}b_{i}b_{i+1} & 1\leq i<n-2\\
& \\
b_{i}b_{j}=b_{j}b_{i} & \left\vert i-j\right\vert \geq2
\end{array}
\right.
\end{eqnarray}
where the notation $b_{i}\equiv b_{i,i+1}$ is used, $b_{i,i+1}$
represents $1_{1}\otimes 1_{2}\otimes 1_{3}\cdots \otimes
S_{i,i+1}\otimes \cdots \otimes 1_{n}$ , and $1_{j}$ is the unit
matrix of the j-th particle.

As is known, Hecke algebras are intimately connected with braiding
groups. In fact, braid algebra is subalgebra of Hecke algebra. And
we can construct a representation of braid algebra from Hecke
algebra. A unitary solution of YBE can also be constructed from a
representation of Hecke algebra. Let us review
Yang-Baxterization\cite{vj,gxw,sun} of Hecke algebra. $M_{i}$, a
Hermitian matrix($\emph{i.e.} M^{\dag}_{i}=M_{i} $), satisfies the
Hecke algebraic relations:
$M_{i}M_{i+1}M_{i}+gM_{i}=M_{i+1}M_{i}M_{i+1}+gM_{i+1}$ and
$M_{i}^{2}=\alpha M_{i}+\beta I_{i}$. For convenience, we set
$\alpha=1$ and $\beta=g=2$. let the unitary Yang-Baxter matrix take
the form,
\begin{equation}\label{ybzation}
\breve{R}_{i}(x)=\rho(x)[\mathbf{1}_{i}+F(x)M_{i}]
\end{equation}
Substituting Eq(\ref{ybzation}) into Eq(\ref{ybe}), one has
$F(x)+F(y)+F(x)F(y)=[1+2F(x)F(y)]F(xy)$. The unitary condition
$(\emph{i.e.},\breve{R}^{\dag}_{i}(x)=\breve{R}^{-1}_{i}(x)=\breve{R}_{i}(x^{-1}))$
can be tenable only on condition that
$F(x)+F(x^{-1})+F(x)F(x^{-1})=0$ and
$\rho(x)\rho(x^{-1})[1+2F(x)F(x^{-1})]=0$. In addition, the initial
condition $\breve{R}_{i}(x=1)=I_{i}$ yields F(x=1)=0 and
$\rho(x=1)=1$. Taking account into these conditions, we obtain a set
solutions of F(x) and $\rho(x)$,
$$\rho(x)=\frac{2x+x^{-1}}{3},~~~F(x)=-\frac{x-x^{-1}}{2x+x^{-1}}.$$
In this paper, we choose basis $\{|11\rangle,|10\rangle,|01\rangle,|1-1\rangle,|00\rangle,|-11%
\rangle,|0-1\rangle,|-10\rangle,|-1-1\rangle\}$ as the standard
basis. Based on calculation, a $9\times 9$ matrix M which satisfies
the Hecke algebraic relations is realized as,
\begin{eqnarray}\label{malgebra}
\begin{array}{lll}
M^{ab}_{cd}=q_{1}\delta_{ab1}|_{a\neq c\neq d}+q_{2}\delta_{ab0}|_{a\neq c\neq d}\\
\\~~~~~~~~+ Q^{-1}\delta_{ab-1}|_{a\neq c\neq d}+q_{1}^{-1}\delta_{cd1}|_{c\neq a\neq b}\\
   \\~~~~~~~~+q_{2}^{-1}\delta_{cd0}|_{c\neq a\neq b}+Q\delta_{cd-1}|_{c\neq a\neq b}\\
   \\~~~~~~~~+\delta_{ad}\delta_{bc}\mid_{a\neq b}
\end{array}
\end{eqnarray}
Where $q_{1}=e^{i\varphi_{1}}$, $q_{2}=e^{i\varphi_{2}}$ and
$Q=q_{1}q_{2}$, with the parameters $ \varphi_{1}$ and $\varphi_{2}$
both are real. The denotes $M^{ab}_{cd}\equiv M_{ab,cd}$ are used.
The denote $\delta_{abc}=1$, if and only if $a=b=c$; otherwise, the
denote $\delta_{abc}=0$. This solution is not equivalent to the
solution in Ref.\cite{sun}. Substituting Eq(\ref{malgebra}) into
Eq(\ref{ybzation}), the unitary solution of YBE can be obtained as
following,
\begin{equation}\label{solution}
   \breve{R}(x,\varphi_{1},\varphi_{2})^{ab}_{cd}=\rho(x)[\delta_{abcd}+F(x)M^{ab}_{cd}]
\end{equation}
The matrix form of $\breve{R}_{i}(x,\varphi_{1},\varphi_{2})$ can be
recast as,
\begin{eqnarray}\label{matrixform}
 \check{R}_{i}(x,q_{i})=\frac{1}{3}\left(
  \begin{array}{ccccccccc}
   b & 0 & 0 & 0 & 0 & 0 & aq_{1} &aq_{1} & 0 \\
    0 & b & a & 0 & 0 &0 & 0 & 0 & aQ \\
    0 & a & b & 0 & 0 & 0 & 0 & 0 & aQ \\
    0 & 0 & 0 & b & \frac{a}{q_{2}} & a & 0 & 0 & 0 \\
  0 & 0 & 0 & aq_{2} & b & aq_{2} & 0 & 0 & 0 \\
    0 & 0 & 0 & a & \frac{a}{q_{2}} & b & 0 & 0 & 0 \\
    \frac{a}{q_{1}} & 0 & 0 & 0 & 0 & 0 & b & a & 0 \\
    \frac{a}{q_{1}} & 0 & 0 & 0 & 0 & 0 & a & b & 0 \\
   0 & \frac{a}{Q} & \frac{a}{Q} & 0 & 0 & 0 & 0 & 0 & b \\
  \end{array}
\right)
\end{eqnarray}
where a=$x^{-1}-x$, $b=2x+x^{-1}$.
The Gell-Mann matrices, a basis for the Lie algebra SU(3)\cite{wp}, $\lambda_{u}$ satisfy $%
[I_{\lambda},I_{\mu}]=if_{\lambda\mu\nu}I_{\nu}
(\lambda,\mu,\nu=1,\cdot \cdot \cdot ,8)$, where
$I_{\mu}=\frac{1}{2}\lambda_{\mu}$. For the later
convenience, we denote $I_{\lambda}$ by, $I_{\pm}=I_{1}\pm iI_{2}$, $%
V_{\pm}=V_{4}\mp iV_{5}$,$U_{\pm}=I_{6}\pm iI_{7}$,
$Y=\frac{2}{\sqrt{3}}I_{8} $. In this work, we get rise to three
sets of realization of $SU(3)$ as:
\begin{eqnarray*}\label{su31}
\left\{
\begin{array}{lll}
I_{\pm}^{(1)}=I_{1}^{\pm}I_{2}^{\mp},~~~U_{\pm}^{(1)}=U_{1}^{\pm}V_{2}^{
\mp},~~~V_{\pm}^{(1)}=V_{1}^{\pm}U_{2}^{\mp},\\
&\\
I_{3}^{(1)}=\frac{1}{3}(I_{1}^{3}-I_{2}^{3})+\frac{1}{2}%
(I_{1}^{3}Y_{2}-Y_{1}I_{2}^{3}),\\
&\\
Y^{(1)}=\frac{1}{3}(Y_{1}+Y_{2})-\frac{2}{3}I_{1}^{3}I_{2}^{3}-\frac{1}{2}
Y_{1}Y_{2};
\end{array}
\right.
\end{eqnarray*}
\begin{eqnarray*}\label{su32}
\left\{
\begin{array}{lll}
I_{\pm}^{(2)}=U_{1}^{\pm}U_{2}^{\mp},~~~U_{\pm}^{(2)}=V_{1}^{\pm}I_{2}^{%
\mp},~~~V_{\pm}^{(2)}=I_{1}^{\pm}V_{2}^{\mp} , \\
& \\
I_{3}^{(2)}=\frac{1}{2}[-\frac{1}{3}(I_{1}^{3}-I_{2}^{3})+\frac{1}{2}%
(Y_{1}-Y_{2})+I_{1}^{3}Y_{2}-Y_{1}I_{2}^{3}], \\
&\\
Y^{(2)}=-[\frac{1}{3}(I_{1}^{3}+I_{2}^{3})+\frac{1}{6}(Y_{1}+Y_{2})+\frac{2}{3%
}I_{1}^{3}I_{2}^{3}+\frac{1}{2}Y_{1}Y_{2}] ;&\\
\end{array}
\right.
\end{eqnarray*}
\begin{eqnarray*}\label{su33}
\left\{
\begin{array}{lll}
I_{\pm}^{(3)}=V_{1}^{\pm}V_{2}^{\mp},~~~U_{\pm}^{(3)}=I_{1}^{\pm}U_{2}^{%
\mp},~~~V_{\pm}^{(3)}=U_{1}^{\pm}I_{2}^{\mp}, \\
&\\
I_{3}^{(3)}=\frac{1}{2}[-\frac{1}{3}(I_{1}^{3}-I_{2}^{3})-\frac{1}{2}%
(Y_{1}-Y_{2})+I_{1}^{3}Y_{2}-Y_{1}I_{2}^{3}],\\
&\\
Y^{(3)}=\frac{1}{3}(I_{1}^{3}+I_{2}^{3})-\frac{1}{6}(Y_{1}+Y_{2})-\frac{2}{3}%
I_{1}^{3}I_{2}^{3}-\frac{1}{2}Y_{1}Y_{2}.\\
\end{array}
\right.
\end{eqnarray*}
We denote $I^{(k)}_{\pm}=I^{(k)}_{1}\pm iI^{(k)}_{2}$, $V^{(k)}_{
\pm}=V^{(k)}_{4}\mp iV^{(k)}_{5}$,$U^{(k)}_{\pm}=I^{(k)}_{6}\pm
iI^{(k)}_{7}$, $Y^{(k)}=\frac{2}{\sqrt{3}}I^{(k)}_{8}$$(k=1,2,3)$.
These realizations satisfy the commutation relation
$[I^{(i)}_{\lambda},I^{(j)}_{\mu}]=i\delta_{ij}f_{
\lambda\mu\nu}I^{(i)}_{\nu}$ $(\lambda,\mu,\nu=1,\cdot \cdot \cdot
,8;i,j=1,2,3)$. So the whole tensor space $C^{3}\otimes C^{3}$ is
completely decomposed. In addition, each block of $\breve{R}$-matrix
can be represented by fundamental representation of SU(3) algebra.
i.e. $C^{3}\otimes C^{3}=C^{3}\oplus C^{3}\oplus C^{3}$.

For $i$-th and $(i+1)$-th lattices, $\breve{R}$-matrix can be
expressed in terms of above operators,
\begin{eqnarray*}\label{operator}
\begin{array}{llll}
\breve{R}(\theta,\varphi_{1},\varphi_{2})=\frac{1}{3}a[I_{+}^{(1)}+I_{-}^{(1)}+Q(V_{-}^{(1)}+U_{+}^{(1)})\\
\\~~~~~~~~~~~~~~~~~~+Q^{-1}(U_{-}^{(1)}+ V_{+}^{(1)})+I_{+}^{(2)}+I_{-}^{(2)} \\
\\~~~~~~~~~~~~~~~~~~+q_{1}(V_{+}^{(2)}+U_{-}^{(2)})+q_{1}^{-1}(V_{-}^{(2)}+U_{+}^{(2)}) \\
\\~~~~~~~~~~~~~~~~~~+I_{+}^{(3)}+I_{-}^{(3)} +q_{2}(V_{+}^{(3)}+U_{-}^{(3)})\\
\\~~~~~~~~~~~~~~~~~~+q_{2}^{-1}(V_{-}^{(3)}+U_{+}^{(3)})]+\frac{b}{3}(I\otimes I).
\end{array}
\end{eqnarray*}
We can introduce a new variable with x=$e^{i\theta}$, and $\theta$
may be related with entanglement degree. When one acts
$\breve{R}(\theta,\varphi_{1},\varphi_{2})$ on the separable state
$|mn\rangle$ , he yields the following family of states
$|\psi\rangle_{mn}=\sum_{ij=11}^{-1-1}\breve{R}^{ij}_{mn}|mn\rangle$(m,n=1,0,-1).
For example, if m=1 and n=1,
$|\psi\rangle_{11}=\frac{1}{3}(b|11\rangle+aq_{1}^{-1}|0-1\rangle+
aq_{1}^{-1}|-10\rangle)$.  By means of negativity\cite{kz,gr,jm}, we
study these entangled states. The negativity for two qutrits is
given by,
\begin{equation}
N(\rho)\equiv\frac{\|\rho^{T_{A}}\|-1}{2},
\end{equation}
where $\|\rho^{T_{A}}\|$ denotes the trace norm of $\rho^{T_{A}}$,
and $\rho^{T_{A}}$ denotes the partial transpose of the bipartite
state $\rho$. \emph{i.e.},
$(\rho)^{i_{A}i_{B}}_{j_{A}j_{B}}=(\rho^{T_{A}})^{j_{A}i_{B}}_{i_{A}j_{B}}$.
In fact, $N(\rho)$ corresponds to the absolute value of the sum of
negative eigenvalues of $\rho^{T_{A}}$, and negativity vanishes for
unentangled states \cite{gr}. Then we can obtain the negativity of
the state $|\psi\rangle_{11}$ as
\begin{equation}\label{N}
N(\theta)=\frac{4}{9}(sin^{2}\theta+|\sin\theta|\sqrt{%
1+8cos^{2}\theta}).
\end{equation}
When $|a|=|b|$, namely $x=e^{i\frac{\pi}{3}}$, the state
$|\psi\rangle_{11}$ becomes the maximally entangled state of two
qutrits as $|\psi\rangle_{11}=\frac{1}{\sqrt{3}}%
(e^{i\frac{\pi}{6}}|11\rangle-iq_{1}^{-1}|0-1\rangle-iq_{1}^{-1}|-10\rangle)$.
In general, if one acts the unitary Yang-Baxter matrix
$\breve{R}(x)$ on the
basis $\{|11\rangle,|10\rangle,|01\rangle,|1-1\rangle,|00\rangle,|-11%
\rangle,|0-1\rangle,|-10\rangle,|-1-1\rangle\}$, he will obtain the
same negativity as Eq(\ref{N}). It is easy to check that the
negativity ranges from 0 to 1 when the parameter $\theta$ runs from
0 to $\pi$. But for $\theta \in [0,\pi]$, the negativity is not a
monotonic function of $\theta$. And when $x=e^{i\frac{\pi}{3}}$, he
will generate nine complete and orthogonal maximally entangled
states for two qutrits. The QE doesn't dependent on the parameters
$\varphi_{1}$ and $\varphi_{2}$. So one can verify that parameter
$\varphi_{1}$ and $\varphi_{2}$ may be absorbed into a local
operation. Base on numerical calculation, the universality of YBE is
proved by Chen \emph{et.al.} in Ref.\cite{cxg1}. This unitary
solution of YBE can generate entangled states, this solution may be
a universal quantum gate.
\section{Yang-Baxter Hamiltonian and BP}\label{sec3}
A Hamiltonian of the Yang-Baxter system can be constructed from the
$\breve{R}(\theta , \varphi_{1},\varphi_{2})$-matrix. As shown in
Ref.\cite{ckg}, the Hamiltonian is obtained through the
Schr\"{o}dinger evolution of the entangled states. Let the
parameters $\varphi_{i}$ be time-dependent as
$\varphi_{i}=\omega_{i} t$. The Hamiltonian reads,
\begin{eqnarray}
\hat{H} &=& i\hbar\frac{\partial\breve{R}(\theta ,
\varphi_{1},\varphi_{2})}{\partial t}\breve{R}^{\dag }(\theta , \varphi_{1},\varphi_{2})\nonumber\\
&=& \bigoplus_{k=1}^{3} H^{(k)},
\end{eqnarray}
where the superscript $k$ denotes the $k$-th subsystem. The $k$-th
subsystem's Hamiltonian $H^{(k)}$ can be obtained as following,

\begin{eqnarray}\label{sub1}
\begin{array}{lll}
H^{(1)}=C(1)[\frac{\sqrt{2}}{6}sin\theta (I_{+}^{(1)}+I_{-}^{(1)})+\frac{\sqrt{2}}{2}sin\theta Y^{(1)}\\
\\~~~~-\frac{\sqrt{2}}{12}ib^{*}Q(V_{-}^{(1)}+U_{+}^{(1)})+\frac{\sqrt{2}}{12}ibQ^{-1}(V_{+}^{(1)}+U_{-}^{(1)})]
\end{array}
\end{eqnarray}
\begin{eqnarray}\label{sub2}
\begin{array}{lll}
H^{(2)}=C(2)[-\frac{\sqrt{2}}{6}sin\theta (I_{+}^{(2)}+
I_{-}^{(2)})-\frac{\sqrt{2}}{2}sin\theta Y^{(2)}]\\
\\~~~~~+\frac{\sqrt{2}}{12}ib^{*}q_{1}^{-1}(U_{+}^{(2)}+V_{-}^{(2)})
-\frac{\sqrt{2}}{12}ibq_{1}(V_{+}^{(2)}+U_{-}^{(2)})]
\end{array}
\end{eqnarray}
\begin{eqnarray}\label{sub3}
\begin{array}{lll}
H^{(3)}=C(3)[-\frac{\sqrt{2}}{6}sin\theta (I_{+}^{(3)}+ I_{-}^{(3)})
-\frac{\sqrt{2}}{2}sin\theta Y^{(3)}\\
\\~~~~~+\frac{\sqrt{2}}{12}ib^{*}q_{2}^{-1}(U_{+}^{(3)}+V_{-}^{(3)})
-\frac{\sqrt{2}}{12}ibq_{2}(V_{+}^{(3)}+U_{-}^{(3)})]
\end{array}
\end{eqnarray}
 Where $C(1)=-\frac{4\sqrt{2}\hbar \Omega sin\theta}{3}$,
$C(2)=-\frac{4\sqrt{2}\hbar \omega_{1} sin\theta}{3}$ ,
$C(3)=-\frac{4\sqrt{2}\hbar \omega_{2} sin\theta}{3}$ and
$\Omega\equiv \omega_{1}+\omega_{2}$.  In terms of
$I_{\lambda}^{(k)}$($\lambda=1,2,\cdots,8; k=1,2,3$), the
Hamiltonian can be recast as following,
\begin{eqnarray}\label{H11}
H^{(k)}=C(k)\sum_{\lambda=1}^{8}B_{%
\lambda}^{(k)}I_{\lambda}^{(k)}.
\end{eqnarray}
 Compare
Eq(\ref{sub1}), Eq(\ref{sub2}), Eq(\ref{sub3}) with Eq(\ref{H11}),
one can obtain $B_{\lambda}^{(k)}$ as following,
\begin{eqnarray*}
\left\{
\begin{array}{llllllll}
B_{1}^{(1)} = \frac{\sqrt{2}}{3}sin\theta ~;~B_{2}^{(1)} =B_{3}^{(1)}= 0\\
&\\
B_{4}^{(1)} = -\frac{\sqrt{2}}{6}sin\theta
  \cos\omega(1) t+\frac{\sqrt{2}}{2}cos\theta\sin\omega(1) t\\
&\\
B_{5}^{(1)} = \frac{\sqrt{2}}{6}sin\theta
 \sin\omega(1) t+\frac{\sqrt{2}}{2}cos\theta\cos\omega(1) t\\
&\\
B_{6}^{(1)} = -\frac{\sqrt{2}}{6}sin\theta
  \cos\omega(1) t+\frac{\sqrt{2}}{2}cos\theta\sin\omega(1) t\\
&\\
B_{7}^{(1)} = \frac{\sqrt{2}}{6}sin\theta
 \sin\omega(1) t+\frac{\sqrt{2}}{2}cos\theta\cos\omega(1) t\\
&\\
B_{8}^{(1)}=\frac{\sqrt{2}}{2}sin\theta
\end{array}
\right.
\end{eqnarray*}
\begin{eqnarray*}
\left\{
\begin{array}{llllllll}
B_{1}^{(i)} = -\frac{\sqrt{2}}{3}sin\theta ~;~B_{2}^{(i)} =B_{3}^{(i)}=0\\
&\\
B_{4}^{(i)} = \frac{\sqrt{2}}{6}sin\theta
  \cos\omega(i)t+\frac{\sqrt{2}}{2}cos\theta\sin\omega(i)t\\
&\\
B_{5}^{(i)} = \frac{\sqrt{2}}{6}sin\theta
  \sin\omega(i)t-\frac{\sqrt{2}}{2}cos\theta\cos\omega(i)t\\
&\\
B_{6}^{(i)} = \frac{\sqrt{2}}{6}sin\theta
  \cos\omega(i)t+\frac{\sqrt{2}}{2}cos\theta\sin\omega(i)t\\
&\\
B_{7}^{(i)} = \frac{\sqrt{2}}{6}sin\theta
  \sin\omega(i)t-\frac{\sqrt{2}}{2}cos\theta\cos\omega(i)t\\
&\\
B_{8}^{(i)}=-\frac{\sqrt{2}}{2}sin\theta
\end{array}
\right.
\end{eqnarray*}
Where i=2,3. The denotes $\omega(1)\equiv \Omega$,
$\omega(2)\equiv\omega_{1}$ and $\omega(3)\equiv\omega_{2}$ are
used. The Hamiltonian for the $k$-th subsystem,
$H(\textbf{B}(t)^{(k)})^{(k)}$, depends on the parameters
$B_{\lambda}^{(k)}$($\lambda=1,2,\cdots,8;$), which are the
components of a vector $\textbf{B}^{(k)}$. And $\textbf{B}^{(k)}$
are a set of time-varying parameters controlling the $k$-th
subsystem's Hamiltonian. After time $T^{(k)}$, Hamiltonian returns
to its original form, \emph{i.e.}
$H(\textbf{B}(0))^{(k)}=H(\textbf{B}(T^{(k)}))^{^{(k)}}$. According
to this, one can easily verify periods of the subsystems are
$T^{(1)}=2\pi/\Omega$, $T^{(2)}=2\pi/\omega_{1}$ and
$T^{(3)}=2\pi/\omega_{2}$. The eigenstates of the first subsystem
are found to be,
\begin{eqnarray*}\label{states1}
\begin{array}{lll}
  |E^{(1)}_{+}\rangle =N_{+}^{(1)}[f_{+}^{(1)}(|10\rangle+|01\rangle) +e^{-i\Omega t}|-1-1\rangle]\\
  \\
  |E^{(1)}_{0}\rangle = \frac{1}{\sqrt{2}}(-|10\rangle+|01\rangle) \\
  \\
  |E^{(1)}_{-}\rangle = N_{-}^{(1)}(f_{-}^{(1)}(|10\rangle+ |01\rangle)+e^{-i\Omega t}|-1-1\rangle).
\end{array}
\end{eqnarray*}
with the corresponding eigenvalues $E^{(1)}_{+} = \frac{2\sqrt{2}}{3}%
\hbar\Omega\sin\theta$, $E^{(1)}_{0} = 0$ and $E^{(1)}_{-} = -\frac{2\sqrt{2}%
}{3}\hbar\Omega\sin\theta$. Where\\
$N_{\pm}^{(1)}=\sqrt{\frac{3\pm2\sqrt{2}sin \theta}{6}}$ and
$f_{\pm}^{(1)}=\frac{4sin \theta\mp3\sqrt{2}}{2ib(\theta)}$. For the
second and the third subsystems, the eigenstates are found to be,
\begin{eqnarray*}\label{states2}
\begin{array}{lll}
  |E^{(2)}_{+}\rangle = N_{+}^{(2)}[f_{+}^{(2)}|11\rangle +e^{-i\omega_{1}t}(|0-1\rangle+|-10\rangle)]\\
  \\
  |E^{(2)}_{0}\rangle = \frac{1}{\sqrt{2}}(-|0-1\rangle+|-10\rangle) \\
  \\
  |E^{(2)}_{-}\rangle =N_{-}^{(2)}[f_{-}^{(2)}|11\rangle+ e^{-i\omega_{1}t}(|0-1\rangle+|-10\rangle)].
\end{array}
\end{eqnarray*}
\begin{eqnarray*}\label{states3}
\begin{array}{lll}
  |E^{(3)}_{+}\rangle =N_{+}^{(3)}[f_{+}^{(3)}|00\rangle +e^{-i\omega_{2}t}(|1-1\rangle+|-11\rangle)]\\
  \\
  |E^{(3)}_{0}\rangle = \frac{1}{\sqrt{2}}(-|1-1\rangle+|-11\rangle) \\
  \\
  |E^{(3)}_{-}\rangle = N_{-}^{(3)}[f_{-}^{(3)}|00\rangle+ e^{-i\omega_{2}t}(|1-1\rangle+|-11\rangle)].
\end{array}
\end{eqnarray*}
with the corresponding eigenvalues $E^{(i)}_{+} =
\frac{2\sqrt{2}}{3} \hbar\omega(i)\sin\theta$, $E^{(i)}_{0} = 0$ and
$E^{(i)}_{-} = -\frac{2\sqrt{2} }{3}\hbar\omega(i)\sin\theta$. Where
$N_{\pm}^{(i)}=\sqrt{\frac{3\pm2\sqrt{2}sin \theta}{12}}$ and
$f_{\pm}^{(i)}=\frac{4sin \theta\mp3\sqrt{2}}{ib^{*}(\theta)}$.
According to the definition of the BP \cite{berry}, when the
parameter $\textbf{B}^{(k)}$ is slowly changed around a circuit on
the sphere of direction, then at the end of circuit, the eigenstates
$|E_{\alpha}^{k}\rangle$($\alpha=+,0,-$ )evolves adiabatically from
0 to $T^{(k)}$, the BP accumulated by the states
$|E_{\alpha}^{k}\rangle$ are,

\begin{equation}
\gamma _{\alpha}^{(k)}=i\int_{0}^{T^{(k)}}\langle E_{\alpha }^{(k)}|
\frac{\partial}{\partial t}|E_{\alpha }^{(k)}\rangle dt.
\label{berryphase}
\end{equation}
Substitute these eigenstates into Eq(\ref{berryphase}), one can
obtain the BP for these eigenstates(all phases are defined modulo
2$\pi$ throughout this paper),
\begin{eqnarray}\label{phase}
\left\{
\begin{array}{lll}
\gamma^{k}_{+} &=& -(\frac{1}{2}-\frac{\sqrt{2}}{3}sin \theta)2\pi \\
&\\
\gamma^{k}_{0} &=& 0 \\
&\\
\gamma^{k}_{-} &=& (\frac{1}{2}-\frac{\sqrt{2}}{3}sin \theta)2\pi
\end{array}
\right.
\end{eqnarray}
In fact, the BP of this system can be represented under the
framework of SU(2) algebra. First, we can introduce three sets SU(2)
realizations in terms of three new sets of
operators(Eq(\ref{su31}),Eq(\ref{su32}),Eq(\ref{su33})),
\begin{eqnarray}\label{rlz1}
\left\{
\begin{array}{lll}
S_{+}^{(k)} & = & \frac{1}{\sqrt{2}}(V_{-}^{(k)}+U_{+}^{(k)})\\
&\\
S_{-}^{(k)} & = & \frac{1}{\sqrt{2}}(V_{+}^{(k)}+U_{-}^{(k)})\\
&\\
S_{3}^{(k)} & = &
\frac{3}{4}Y^{(k)}+\frac{1}{4}(I_{+}^{(k)}+I_{-}^{(k)}).
\end{array}
\right.
\end{eqnarray}
They satisfy the algebraic relations of $SU(2)$ group:\\
$[S^{(i)}_{+},S^{(j)}_{-}]=2 \delta_{ij}S^{(i)}_{3}$,
$[S^{(i)}_{3},S^{(j)}_{\pm}]=\pm \delta_{ij}S^{(i)}_{\pm}$,
$(S_{\pm}^{(i)})^{2}=0$($i,j=1,2,3$), with
$S^{(k)}_{\pm}=S^{(k)}_{1}\pm iS^{(k)}_{2}$($k=1,2,3$). By the way,
their second-order Casimir operators are $\mathcal
{J}^{(k)}=\frac{1}{2}(S^{(k)}_{+}S^{(k)}_{-}+S^{(k)}_{-}S^{(k)}_{+})+(S^{(k)}_{3})^{2}$.
One can verify that the eigenvalues of  $\mathcal {J}^{(k)}$ are
$\frac{1}{2}(\frac{1}{2}+1)=\frac{3}{4}$ and $0(0+1)=0$ which
correspond to spin-$1/2$ system and spin-0 system.

When one substitutes these realizations into
Eq(\ref{sub1})-Eq(\ref{sub3}), he can recast Hamiltonian of the
subsystems in terms of SU(2),
\begin{eqnarray}\label{su2hamil}
\begin{array}{lll}
  H^{(1)}=C(1)[\frac{1}{2}(B_{-}^{(1)}S_{+}^{(1)}+B_{+}^{(1)} S_{-}^{(1)})+B_{3}^{(1)} S_{3}^{(1)}]\\
  \\
  H^{(2)}=C(2)[\frac{1}{2}(B_{-}^{(2)}S_{+}^{(2)}+B_{+}^{(2)} S_{-}^{(2)})+B_{3}^{(2)} S_{3}^{(2)}]\\
  \\
  H^{(3)}=C(3)[\frac{1}{2}(B_{-}^{(3)}S_{+}^{(1)}+B_{+}^{(3)} S_{-}^{(3)})+B_{3}^{(3)} S_{3}^{(3)}].
\end{array}
\end{eqnarray}
Where $B_{-}^{(1)}=(B_{+}^{(1)})^{*}=-\frac{1}{3}ib^{*}e^{i\Omega t}
$, $B_{3}^{(1)}=\frac{2\sqrt{2}}{3}\sin\theta$,
$B_{-}^{(i)}=(B_{+}^{(i)})^{*}=\frac{1}{3}ib^{*}e^{-i\omega(i)t}$
and $B_{3}^{(i)}=-\frac{2\sqrt{2}}{3}\sin\theta$(i=1;2). $\omega(i)$
and C(3) are defined below Eq(\ref{H11}). So we can say the whole
system equivalent to three spin-$\frac{1}{2}$ subsystems and three
spin-0 subsystems. In fact, we can introduce a time-independent
$9\times9$ orthogonal matrix O (see Appendix A). By means of O, the
whole system's Hamiltonian $\hat{H}$ and Casimir operators $\mathcal
{J}^{(k)}$ are transformed into block-diagonal matrices. \emph{i.e.}
$\tilde{\hat{H}}=O\hat{H}O^{T}$ and $\tilde{\mathcal
{J}}^{(k)}=O\mathcal {J}^{(k)}O^{T}$ are block-diagonal matrices,
where $O^{T}$ denotes the transpose of matrix $O$. For the subsystem
1, from  Eq(\ref{12}) we can get its Hamiltonian
$\tilde{H}^{(1)}=H_{\frac{1}{2}}^{(1)}\oplus H_{0}^{(1)}$. For
$H_{0}^{(1)}$, the eigenvalue of Casimir operator $\mathcal
{J}^{(1)}$ is 0, and the BP is 0. So we can say the subsystem
Hamiltonian $H_{0}^{(1)}$ is equivalent to a spin-$0$ subsystem. For
$H_{\frac{1}{2}}^{(1)}$, one can introduce two transformations,
$cos\alpha=\frac{2\sqrt{2}}{3}sin\theta$ and
$cos\beta=\frac{-sin\theta cos\Omega t+3cos\theta sin\Omega
t}{\sqrt{9-8\sin^{2}\theta}}$, with $\alpha \in
(arccos\frac{2\sqrt{2}}{3},arccos-\frac{2\sqrt{2}}{3})$ and $\beta
\in[0,2\pi]$. $\alpha$ is time-independent, and $\beta$ is
time-dependent. By means of this transformation, the Hamiltonian
$H^{(1)}_{\frac{1}{2}}$ can be recast as
$H^{(1)}_{\frac{1}{2}}=C(1)(sin\alpha cos\beta S_{1}+sin\alpha
sin\beta S_{2}+cos \alpha S_{3})$. We substitute these
transformations into Eq(\ref{phase}), the BP of the subsystem
$H_{\frac{1}{2}}^{(1)}$ can be recast as,
\begin{eqnarray}
\mathcal{\gamma}_{\pm}^{(1)}=\mp\pi(1-cos\alpha)=\mp\Omega(C)/2
\end{eqnarray}
where $\Omega(C) = 2\pi (1- \cos \alpha)$ is the familiar solid
angle enclosed by the loop on the Bloch sphere, and the parameter
$\alpha$ comes from $\theta$ which comes from the Yang-Baxterization
of the unitary braiding operator. So the BP dependent on spectral
parameter. Under the new basis Eq(\ref{11}), the eigenstates
$|E^{(1)}_{\pm}\rangle$ can be recast as following (we neglected the
global phase factor),
\begin{eqnarray}
|E^{(1)}_{+}\rangle &=&
-e^{-i\beta}\sin\frac{\alpha}{2}|1\rangle+\cos\frac{\alpha}{2}|2\rangle,\\
|E^{(1)}_{-}\rangle &=&
\cos\frac{\alpha}{2}|1\rangle+e^{i\beta}\sin\frac{\alpha}{2}|2\rangle,
\end{eqnarray}
where $|1\rangle=O\frac{1}{\sqrt{2}}(|10\rangle+|01\rangle)$ and
$|2\rangle=O|-1-1\rangle$. As is known to all, they are spin
coherent states. By means of Eq(\ref{10}), the states
$|E^{(1)}_{\pm}\rangle$ can be recast as following,
\begin{eqnarray}\label{coherent}
    |E^{(1)}_{+}\rangle &=& exp[\zeta \tilde{S}_{+}^{(1)}-\zeta^{*}
    \tilde{S}_{-}^{(1)}]|2\rangle,\nonumber\\
    |E^{(1)}_{-}\rangle &=& exp[\zeta \tilde{S}_{+}^{(1)}-\zeta^{*}
    \tilde{S}_{-}^{(1)}]|1\rangle,
\end{eqnarray}
Where $\zeta=e^{-i\beta}\alpha/2$. BP for spin coherent states has
been investigated in Ref\cite{smv}. So we can say the subsystem
$H_{\frac{1}{2}}^{(1)}$ is equivalent to a spin-$\frac{1}{2}$
subsystem. By means of the same method, the Berry phases for
subsystem 2 and 3 may be obtained,
$\mathcal{\gamma}_{\pm}^{(k)}=\mp\pi(1-cos\alpha)=\mp\Omega(C)/2$
and $\mathcal{\gamma}_{0}^{(k)}=0$. The whole system is equivalent
to three spin$-1/2$ subsystems and three spin-0 subsystems. This
Yang-Baxter Hamiltonian system is equivalent to the Hamiltonian in
Ref.\cite{sun}.

\section{Summary}\label{sec5}
In this paper, we have presented a $9 \times 9$ $M$-matrix which
satisfies the Hecke algebraic relations and derived a unitary
$\breve{R}(\theta,\varphi_{1},\varphi_{2})$-matrix via
Yang-Baxterization of the $M$-matrix. In the following, we show that
the arbitrary degree of entanglement for two-qutrit entangled states
can be generated via the unitary
$\breve{R}(\theta,\varphi_{1},\varphi_{2})$ matrix acting on the
standard basis. Then the evolution of the Yang-Baxter system is
explored by constructing a Hamiltonian from the unitary
$\breve{R}(\theta,\varphi_{1},\varphi_{2})$-matrix. In addition, the
BP of the system is investigated.  By means of decomposition of the
tensor product, the Berry phase of the whole system is explained.
The whole system is equivalent to three spin$-\frac{1}{2}$
subsystems and three spin-0 subsystems. Berry phase of this system
is represented under the framework of SU(2) algebra.

 Thermal entanglement in multi-body system is an interesting and nature type of
QE, so it is a good challenge to study the thermal entanglement in
multi-body system. Yang-Baxter Equation is an important tool in this
domain.

\section*{Acknowledgments}
 This work was supported by NSF of
China (Grant No. 10875026).

\begin{appendix}
\section{Block-Diagonalize $\hat{H}$ and $\mathcal {J}^{(k)}$ }
\label{appendixa}

The time-independent $9\times9$ orthogonal matrix O reads,
\begin{eqnarray*}
O=\left(
    \begin{array}{ccccccccc}
      0 & \frac{1}{\sqrt{2}} & \frac{1}{\sqrt{2}} & 0 & 0 & 0 & 0 & 0 & 0 \\
      0 & 0 & 0 & 0 & 0 & 0 & 0 & 0 & 1 \\
      0 & -\frac{1}{\sqrt{2}} & \frac{1}{\sqrt{2}} & 0 & 0 & 0 & 0 & 0 & 0 \\
      1 & 0 & 0 & 0 & 0 & 0 & 0 & 0 & 0 \\
      0 & 0 & 0 & 0 & 0 & 0 & \frac{1}{\sqrt{2}} & \frac{1}{\sqrt{2}} & 0 \\
      0 & 0 & 0 & 0 & 0 & 0 & -\frac{1}{\sqrt{2}} & \frac{1}{\sqrt{2}} & 0 \\
      0 & 0 & 0 & 0 & 1 & 0 & 0 & 0 & 0 \\
      0 & 0 & 0 & \frac{1}{\sqrt{2}} & 0 & \frac{1}{\sqrt{2}} & 0 & 0 & 0 \\
      0 & 0 & 0 & -\frac{1}{\sqrt{2}} & 0 & \frac{1}{\sqrt{2}} & 0 & 0 & 0 \\
    \end{array}
  \right)
\end{eqnarray*}
The orthogonal matrix $O$ satisfies the relation
$OO^{T}=O^{T}O=I_{9\times9}$, where $O^{T}$ denotes the transpose of
matrix $O$.

The orthogonal matrix $O$ transforms the standard basis
$\{|11\rangle, |10\rangle, |01\rangle, |1-1\rangle, |00\rangle,
|-11\rangle, |0-1\rangle, |-10\rangle, |-1-1\rangle\}$ into a new
set of basis. The relations of new basis and old basis are,
\begin{eqnarray*}\label{11}
\left\{
\begin{array}{lll}
|1\rangle=O\frac{1}{\sqrt{2}}(|10\rangle+|01\rangle)\\
& \\
|2\rangle=O|-1-1\rangle\\
&\\
|3\rangle=O\frac{1}{\sqrt{2}}(-|10\rangle+|01\rangle)
\end{array}
\right.
\end{eqnarray*}
\begin{eqnarray*}
 \left\{
\begin{array}{lll}
|4\rangle=O|11\rangle\\
& \\
|5\rangle=O\frac{1}{\sqrt{2}}(|0-1\rangle+|-10\rangle)\\
&\\
|6\rangle=O\frac{1}{\sqrt{2}}(-|0-1\rangle+|-10\rangle)\\
\end{array}
\right.
\end{eqnarray*}
\begin{equation}\label{12}
\left\{
\begin{array}{lll}
|7\rangle=O|11\rangle\\
& \\
|8\rangle=O\frac{1}{\sqrt{2}}(|1-1\rangle+|-11\rangle)\\
&\\
|9\rangle=O\frac{1}{\sqrt{2}}(-|1-1\rangle+|-11\rangle)
\end{array}
\right.
\end{equation}
$\{|1\rangle,|2\rangle,|3\rangle,|4\rangle,|5\rangle,|6\rangle,|7\rangle,|8\rangle,|9\rangle\}$
are a set of new basis.  By means of this set basis, the Hamiltonian
$\hat{H}$ can be recast as block-diagonally form,
\begin{eqnarray}\label{12}
\tilde{\hat{H}}&=&O\hat{H}O^{T}\nonumber\\
&=&diag\{H^{(1)}_{\frac{1}{2}},H^{(1)}_{0},H^{(2)}_{\frac{1}{2}},H^{(2)}_{0},H^{(3)}_{\frac{1}{2}},H^{(3)}_{0}\}\\
&=&\bigoplus_{k=1}^{3} \tilde{H}^{(k)},\nonumber
\end{eqnarray}
where $\tilde{H}^{(k)}=H^{(k)}_{\frac{1}{2}}\oplus H^{(k)}_{0}$,
$H^{(k)}_{\frac{1}{2}}$'s are $2\times2$ matrix, and $H^{(k)}_{0}$
are $1\times1$ matrix with $H^{(k)}_{0}=(0)$. Under the new basis,
the $(3\times3)$-dimension matrix is decomposed into six blocks.

Three sets of $SU(2)$ realizations (\ref{rlz1})can be recast as,
\begin{eqnarray}\label{10}
\tilde{S}^{(1)}_{+}=|1\rangle\langle 2|,~~
\tilde{S}^{(1)}_{-}=|2\rangle\langle 1|,\nonumber\\
\tilde{S}^{(1)}_{3}=\frac{1}{2}(|1\rangle\langle 1|-|2\rangle\langle
2|);\\
\tilde{S}^{(2)}_{+}=|4\rangle\langle 5|,~~
\tilde{S}^{(2)}_{-}=|5\rangle\langle 4|,\nonumber\\
\tilde{S}^{(2)}_{3}=\frac{1}{2}(|4\rangle\langle 4|-|5\rangle\langle
5|);\\
\tilde{S}^{(3)}_{+}=|7\rangle\langle 8|,~~
\tilde{S}^{(2)}_{-}=|8\rangle\langle 7|,\nonumber\\
\tilde{S}^{(2)}_{3}=\frac{1}{2}(|7\rangle\langle 7|-|8\rangle\langle
8|).
\end{eqnarray}
The seconde-order Casimir operators are $\tilde{\mathcal
{J}}^{(1)}=\frac{3}{4}(|1\rangle\langle 1|+|2\rangle\langle 2|)$,
$\tilde{\mathcal {J}}^{(2)}=\frac{3}{4}(|4\rangle\langle
4|+|5\rangle\langle 5|)$, $\tilde{\mathcal
{J}}^{(3)}=\frac{3}{4}(|7\rangle\langle 7|+|8\rangle\langle 8|)$.
\end{appendix}

\end{document}